\documentclass[aps,pra,twocolumn,superscriptaddress,showpacs,floatfix,a4paper]{revtex4}
\usepackage{graphicx,epsfig,bm}

\begin{document}

\title{Perfect State Transfer, Effective Gates and Entanglement Generation in
Engineered Bosonic and Fermionic Networks}

\date{\today}

\author{Man-Hong Yung}
\email[email: ]{myung2@uiuc.edu}

\affiliation{Physics Department, University of Illinois at
Urbana-Champaign, Urbana IL 61801-3080, USA}

\affiliation{Physics Department, The Chinese University of Hong
Kong, Hong Kong}

\author{Sougato Bose}
\email[email: ]{sougato.bose@qubit.org}

\affiliation{Department of Physics and Astronomy, University
College London, Gower Street, London WC1E 6BT}

\pacs{03.67.Hk, 03.67.Mn, 03.67.Lx}

\begin{abstract}
We show how to achieve perfect quantum state transfer and construct
effective two-qubit gates between distant sites in engineered
bosonic and fermionic networks. The Hamiltonian for the system can
be determined by choosing an eigenvalue spectrum satisfying a
certain condition, which is shown to be both sufficient and
necessary in mirror-symmetrical networks. The natures of the
effective two-qubit gates depend on the exchange symmetry for
fermions and bosons. For fermionic networks, the gates are
entangling (and thus universal for quantum computation). For bosonic
networks, though the gates are not entangling, they allow two-way
simultaneous communications. Protocols of entanglement generation in
both bosonic and fermionic engineered networks are discussed.
\end{abstract}

\maketitle

\section{Introduction}

Quantum state transfer, two-qubit gates, and entanglement are
essential in quantum information theory and quantum computation
\cite{Nielsen00}. Recently, there have been many proposals
\cite{{Bose02},{Christandl04},{Yung03},{Osborne04},{Albanese04},{Clark04},{Burgarth04},{Plenio02},{Benjamin04},{Amico04},{Biswas03}}
exploiting the \emph{free} evolution of spin networks for
accomplishing these tasks. The main idea is to minimize the spatial
and dynamical control, which is experimentally challenging, on the
interactions between qubits. Imperfect state transfer over
homogeneous spin chains has been studied \cite{Bose02,{Yung03}} for
Heisenberg and $XY$ Hamiltonians. A measurement-based state transfer
scheme \cite{Burgarth04} has been suggested for dual-spin channels.
Perfect state transfer \cite{Christandl04}, state inversion
\cite{Albanese04}, and graph state generation \cite{Clark04} have
been proposed for engineered spin chains in which the couplings
between qubits are tunable. Quantum computation using permanently
coupled spin chains has been proposed \cite{{Yung03},Benjamin04}.
Furthermore, other dynamical properties \cite{Amico04} of spin
chains and state transfer schemes \cite{Biswas03} have also been
studied.

In this paper, we generalize the results in
\cite{{Christandl04},Yung03,{Albanese04}} for engineered networks.
In Refs. \cite{Christandl04} and \cite{Albanese04} two types of
engineered networks which accomplish perfect quantum state transfer
have been presented. These networks depend on the known properties
of special functions, and hence the choice of the eigenvalue
spectrum and the network couplings is limited. One of the aims of
this paper is to show how one could ``design'' such engineered
networks without reference to any special functions. One simply has
to choose an eigenvalue spectrum from an infinite set of
possibilities satisfying a certain condition [cf. Eq.
(\ref{eq:spectrum})], which is both sufficient and necessary. The
network couplings can then be found by solving a structured inverse
eigenvalue problem. As a consequence of this approach we note that
even a single infinitely deep square well or a single harmonic well
enables perfect state transfer from across a distance. Inspired by
the recent rapid experimental development in optical lattices (see,
e.g., \cite{Duan03} and references therein), our discussion will be
presented in terms of fermionic and bosonic networks with the
presence or absence of a boson or fermion at a site representing the
$0$ and $1$ states of a qubit. Certain spin networks are classified
to be fermionic, as we shall discuss.

Another aim of this paper is to show that effective two-qubit gates
over remote qubit pairs can be constructed in those engineered
networks. For fermionic ones, including spin chains, the effective
gates are entangling and hence universal for quantum computation.
For bosonic networks, the gates are not entangling, but they allow
two-way communication for different pairs of sites simultaneously
without mutual interference. Finally, protocols for entanglement
generation and transfer will also be discussed. In contrast to the
scheme proposed in \cite{Khaneja02}, these protocols require minimal
spatial and temporal control on individual qubits.

\section{Engineered Networks}

We start with a system consisting of spinless fermions (or bosons)
hopping freely in a network of $N$ lattice sites. In fact, the
particles need not literally be spinless, but they all need to be
polarized in the same spin state, and there should not be any
interactions involving spin. The Hamiltonian is therefore of the
following form:
\begin{equation}\label{eq:Hamiltonian}
H = \sum\limits_{\left\langle {i,j} \right\rangle } {} \omega _{i j}
\left( {a_i^\dag  a_j  + a{_j^\dag} a_i    } \right) +
 \sum\limits_{j = 1}^N { \lambda_j \, n_j } \quad ,
\end{equation}
where ${\left\langle {i,j} \right\rangle }$ denotes nearest-neighbor
coupling, $\omega_{ij}$ is the time-independent coupling constant
between the site $i$ and site $j$, and $\lambda_j$ represents the
strength of the external static potential at site $j$. The
annihilation operators $a_j$ obey the standard (anti)commutation
relations for bosons (fermions) and $n_j=a^\dag_j a_j$ is the number
operator. This model may be considered as the strong tunneling limit
of the Hubbard model \cite{Zanardi02} for fermions and Bose-Hubbard
model \cite{Sachdev99} for bosons. In particular, for
one-dimensional fermionic chains, this model can be mapped to spin
chains in which spins are coupled through the $XY$ Hamiltonian
\begin{equation}
H = \frac{1}{2}\sum\limits_{j = 1}^{N-1} {} \omega _{j} \left(
{\sigma^x_j \sigma^x_{j+1}  + \sigma^y_j \sigma_{j+1}^y } \right)+
\frac{1}{2}\sum\limits_{j = 1}^N {} \lambda _j \left( {\sigma _j^z
+ 1} \right)
\end{equation}
by the Jordan-Wigner transformation \cite{Jordan28}. Therefore,
such spin chains will be classified as fermionic, even though the
individual spins are distinguishable.

Since the Hamiltonian $H$ commutes with the total number operator
$n_{tot}={\sum\nolimits_{j = 1}^{N}{n_j } }$ or the total $z$-spin
operator $S^z_{tot}={\sum\nolimits_{j = 1}^N {\sigma^z_j } }$, the
Hilbert space can be decomposed into subspaces consisting of the
eigenstates of $n_{tot}$ or $S^z_{tot}$. Furthermore, as the
particles are noninteracting, the eigenstates in the $n$-particle
subspace are the antisymmetrized (symmetrized) products of the
single-particle eigenstates for fermions (bosons).

\subsection{Quantum State Transfer}

Quantum state transfer over a network is similar to the quantum
random walk problem, where a variety of networks are equivalent to
one-dimensional chains \cite{Farhi98,{Christandl04}}. Therefore, we
will now focus on a chain of $N$ sites. For $j=1,2,...,N$, let
$\left| \bm j \right\rangle$ be the state where a single fermion (or
boson) is at the site $j$ but is in the empty state $\left| 0
\right\rangle$ for all other sites and $\left| \bm 0 \right\rangle$
be the vacuum state where all sites are empty. For spin chains,
$\left| \bm 0 \right\rangle$ corresponds to the state where all the
spins are in the spin-down state $\left|  \downarrow \right\rangle$
and $\left| \bm j \right\rangle$ corresponds to a spin-up state
$\left|  \uparrow \right\rangle$ for the $j$th spin and spin-down
for all other spins. The Hamiltonian in this single-particle
subspace can be written in a tridiagonal form, which is real and
symmetric:
\begin{equation}\label{eq:Tridiagonal}
H_N = \left( {\begin{array}{*{20}c}
   \lambda_1  & {\omega _{1} } & 0 &  \cdots  & 0  \\
   {\omega _{1} } & \lambda_2 & {\omega _{2} } &  \cdots  & 0  \\
   0 & {\omega _{2} } & \lambda_3 &  \cdots  & 0  \\
    \vdots  &  \vdots  &  \vdots  &  \ddots  & {\omega _{{N{-}1}} }  \\
   0 & 0 & 0 & {\omega _{N{-}1 } } & \lambda_N  \\
\end{array}} \right) \quad .
\end{equation}

The quantum state transfer protocol involves two steps:
initialization and evolution. First, a quantum state $\alpha \left|
0 \right\rangle  + \beta \left| 1 \right\rangle $ to be sent is
encoded at site $x$. The initial state of the network is described
by $\left| \bm {\varphi}_x \right\rangle  = \alpha \left| \bm 0
\right\rangle  + \beta \left| \bm x \right\rangle$. Then, the
network couplings $\omega_j$ and $\lambda_j$ are switched on and the
whole system is allowed to evolve under $U\left( t \right) = \exp
\left( { - iHt} \right)$ for a fixed time interval $t=\tau$. The
final state becomes
\begin{equation}
U(\tau)\left| {\bm \varphi _x } \right\rangle  = \alpha \left| \bm
0 \right\rangle  + \beta \sum\limits_{j = 1}^N {} f_{j,x}^N
(\tau)\left| \bm j \right\rangle \quad,
\end{equation}
where $f_{j,x}^N ( \tau ) = \left\langle \bm j \right|e^{ - iH\tau
} \left| \bm x \right\rangle$. Any site $y$ is in a mixed state if
$\left| {f_{y,x}^N ( \tau  )} \right| < 1$, which also implies
that the state transfer from site $x$ to $y$ is imperfect. Our
goal here is to find a set of $\omega_j$ and $\lambda_j$ to
realize perfect state transfer.

In \cite{Albanese04}, it is shown that when the couplings are chosen
such that $H_N=S_x$, where $S_x$ is the $x$ component of the spin
operator of a spin $S=(N{-}1)/2$, or alternatively $H_N= \mathbf{L}
\cdot \mathbf{S}$, subject to the constraint that the $z$ component
of the total angular momentum $\mathbf{J}=\mathbf{L}+\mathbf{S}$
being zero, then a mirror inversion of eigenstates with respect to
the center of a linear chain can be implemented. This implies that a
quantum state at site $x$ can be transferred perfectly to its
\textit{mirror-conjugate} site $\bar x = N{-}x{+}1$,
\begin{equation}\label{eq:state transfer}
U( \tau  )\left| \bm {\varphi}_x \right\rangle  = \alpha \left|
\bm 0 \right\rangle  + e^{-i\phi_N } \beta  \left| \bm {\bar{x}}
\right\rangle \quad ,
\end{equation}
where in general $\phi_N \ne 0$ and a single-qubit operation on the
site $\bar x$ is required to remove it, in order to reconstruct the
original state there.

In the next section, we will introduce a systematic way to find the
sets of $\omega_j$ and $\lambda_j$ for state inversion and hence
perfect state transfer even if we did not use any of the above
examples. Instead of solving the eigenvalue problem, we will first
{\em choose} a desired eigenvalue spectrum [cf. Eq.
(\ref{eq:spectrum})] for $H_N$ and the solutions for $\omega_j$ and
$\lambda_j$ can be found from the spectrum and the symmetrical
properties of $H_N$. It is therefore an inverse eigenvalue problem.

\subsection{Symmetrical properties of $H_N$}

State inversion by free evolution crucially depends on (a)~the
reflection symmetry and (b) the eigenvalue spectrum. By reflection
symmetry, we mean for $j=1,2,...,\left\lfloor {N/2}
\right\rfloor$,
\begin{equation}\label{eq:persymmetric}
\lambda _j  = \lambda _{\bar j} \quad {\rm{and}} \quad \omega _j =
\omega _{N{-}j} \ne 0   \quad .
\end{equation}
Thus, $H_N$ has double symmetries (also known as persymmetric),
along both the main diagonal and the second diagonal. We shall now
show that if the above symmetries are present in $H_N$, then one
only needs the eigenvalue spectrum to satisfy a certain condition
[cf. Eq. (\ref{eq:spectrum})] in order to achieve state inversion.
This condition will later be shown not only sufficient but also
necessary for state transfer in mirror-symmetric networks. As a
consequence of the above symmetries, the eigenvectors $\left| \bm
{e_k } \right\rangle = \sum\nolimits_{j = 1}^N {a_j^k } \left| \bm j
\right\rangle$ have definite parities---i.e. being either even or
odd with respect to the mirror-conjugate operation $j \rightarrow
\bar j$. In fact, the eigenvectors can be determined [cf. Eq.
(\ref{eq:eigenvector})] explicitly. However, we need to know which
one would change sign when inverted. This can be determined by the
interlacing property described below.

Let $P_N(E) = \prod\nolimits_{k = 0}^{N - 1} {\left( {E {-} E_k }
\right)}$ be the characteristic polynomials of $H_N$ and denote the
$j$th leading principal minor (i.e., the characteristic polynomial
obtained by the first $j$ rows and columns of a matrix) of the
matrix $EI - H_N$ by $P_j(E)$, where $I$ is the $N \times N$
identity matrix and $E$ is a real number. With $P_0 \equiv 1$ and
$P_1 \equiv E {-} \lambda_1$, the sequence of $P_j(E)$ is a Sturm
sequence \cite{Wilkinson65} and for $j=1,2,...,N$, it satisfies a
recurrence relation
\begin{equation}
P_j( E) = \left( E {-} {\lambda_j } \right) P_{j-1}(E) - \omega
_{j{-}1}^2 P_{j{-} 2}( E ) \quad .
\end{equation}
The Sturm sequence has an important property: the roots $E^j_k$ of
$P_j$ interlace those of $P_{j-1}$---i.e.
\begin{equation}\label{eq:interlace}
E_{j-1}^{j} < E_{j-2}^{j - 1} < E_{j-2}^{j}  < \cdots <E_1^{j}<
E_{0}^{j {-} 1} < E_0^{j} \quad .
\end{equation}
This implies that
\begin{equation}\label{eq:signs}
{\rm sgn}\left[P_{N-1}(E_k)\right]= (-1) \times {\rm
sgn}\left[P_{N-1}(E_{k-1})\right] \quad ,
\end{equation}
where $E_k\equiv E_k^N$ and $P_N(E_k)=0$. We shall show immediately
that this interlacing property of the Sturm sequence determines the
parity of the eigenvectors.

\subsubsection*{Parity}

It is known that the coefficients $a_j^k$, $j=2,3,...,N$, of the
eigenvectors are given \cite{Wilkinson65} by
\begin{equation}\label{eq:eigenvector}
a_j^k  =  \frac{{P_{j-1} ( {E_k } )}}{{\omega _1 \omega _2 \cdots
\omega _{j-1} }} a_1^k \quad ,
\end{equation}
with $a_1^k$ determined by the normalization condition
$\sum\nolimits_{j = 1}^N {\left| {a_j^k } \right|} ^2  = 1$. We note
that the parity of the eigenvectors can be determined by checking
the relative sign of \emph{any} pair of mirror-conjugate
coefficients. For convenience, we consider ${\rm sgn}\left[a_N^k
/a_1^k\right]={\rm sgn}\left[P_{N-1}(E_k)/\omega _1 \omega _2
\cdots\omega _{N-1}\right]=(-1)^{\nu} \times {\rm
sgn}\left[P_{N-1}(E_k)\right] $, where $(-1)^{\nu}\equiv {\rm
sgn}\left[\omega _1 \omega _2 \cdots\omega _{N-1}\right]$. From Eqs.
(\ref{eq:signs}) and (\ref{eq:eigenvector}), if the eigenvectors are
ordered in decreasing eigenvalues---i.e.,
$E_0>E_1>\cdots>E_{N-1}$--- the parities of them change
\emph{alternatively}. Since $P_{N-1}(E_0)>0$, the parity of the
highest energy eigenstate $\left| \bm {{e}}_0 \right\rangle$ is only
determined by $(-1)^\nu$. It is even (i.e., $\nu=0$), if all
$\omega_j>0$. As the parity changes alternatively, once the parity
of $\left| \bm {{e}}_0 \right\rangle$ is known, the parities of all
other eigenvectors can be inferred immediately. These can be
summarized as
\begin{equation}
\left| \bm {\bar {e}}_k \right\rangle  = \left( { - 1}
\right)^{k+\nu} \left| \bm {e}_k \right\rangle \quad ,
\end{equation}
for $k=0,1,...,N-1$, where  $\left| \bm {\bar e }_k \right\rangle
\equiv \sum\nolimits_{j = 1}^N {} a_j^k \left| \bm {\bar j}
\right\rangle$.

\subsection{Mirror Inversion}

Next, we require, for some time interval $\tau$, the eigenvalue
spectrum of $H_N$ to satisfy the relation
\begin{equation}\label{eq:spectrum}
e^{ - iE_k \tau }  = \left( { - 1} \right)^{\pm k} e^{ - i\phi _N
} \quad ,
\end{equation}
where $\phi_N$ is independent of $k$ and the $\pm$ sign has to be
taken consistently for all $k$. For simplicity, we assume all
$\omega_j>0$. Consider $U( \tau )\left| \bm x \right\rangle =
\sum\nolimits_{k = 0}^{N {-} 1} {} e^{ - i E_k \tau } \left| \bm
{e}_k \right\rangle \! \left\langle \bm {e}_k | \bm x
\right\rangle$. When $\left| \bm {e}_k \right\rangle$ is replaced
with $\left( {- 1} \right)^{-k} \left| \bm {\bar e}_k \right\rangle
$, together with the relation in Eq. (\ref{eq:spectrum}), the
completeness relation $I = \sum\nolimits_{k = 0}^{N - 1} {} \left|
\bm {\bar e}_k \right\rangle \! \left\langle \bm {\bar e}_k
\right|$, and the double inversion relation $\left\langle \bm {e}_k
| \bm j \right\rangle = \left\langle \bm {\bar e}_k | \bm {\bar j}
\right\rangle$, one can show that
\begin{equation}
U(\tau)\left| \bm x \right\rangle  = e^{-i\phi _N } \left| \bm
{\bar x} \right\rangle \quad .
\end{equation}
From Eq. (\ref{eq:state transfer}), consequently, quantum states can
be transported from any site $x$ to its mirror-conjugate site $\bar
x$ after a fixed period~$\tau$. Once a spectrum is determined, the
search for the solutions of $\omega_j$ and $\lambda_j$ becomes an
inverse eigenvalue problem. There are some efficient algorithms
available in the literature for accomplishing the task---for example
\cite{Gladwell86} and references therein.

Here we also note that the condition in Eq. (\ref{eq:spectrum}) is
not only sufficient but also \emph{necessary} for perfect state
transfer in mirror-symmetric networks. To prove that it is
necessary, we set for some time $\tau$, $1 = \left| {\left\langle
\bm {\bar x} \right|U( \tau )\left| \bm x \right\rangle } \right| =
\left| \sum\nolimits_{k}  \left| {\left\langle {{\bm e}_k } |\bm x
\right\rangle } \right|^2 e^{i\varphi _k }\right| \le
\sum\nolimits_k \left| {\left\langle {{\bm e}_k } | {\bm x}
\right\rangle } \right|^2  = 1 $, where $e^{i\varphi _k } \equiv e^{
- iE_k t} \left( { - 1} \right)^k$ and we have used the
normalization condition in the last step. As the above equality must
hold, $e^{i\varphi _k }$ should be a constant phase (independent of
$k$), and hence the condition in Eq. $(\ref{eq:spectrum})$ follows.

\subsubsection*{Example}

Two types of spectrums, $E_k=-k$ and $E_k=k(k+q)$ for some rational
number $q$ and $k=0,1,2,...,N$, suggested in \cite{Albanese04} can
easily be shown to satisfy Eq. (\ref{eq:spectrum}). However, these
spectra are related to some known examples of special functions. To
illustrate the generality of the method, we consider a $4 \times 4$
tridiagonal matrix with the spectrum $E_0=1,E_1=2,E_2=3$, and
$E_3=2(1+m)$, for any integer $m \geq 1$. The condition
(\ref{eq:spectrum}) is satisfied with $\tau=\pi$ and $\phi_N=0$. One
of the solutions for the Hamiltonian [of the form of
Eq.(\ref{eq:Tridiagonal})] is found to be
\begin{equation}
\left( {\begin{array}{*{20}c}
   a & c & 0 & 0  \\
   c & b & d & 0  \\
   0 & d & b & c  \\
   0 & 0 & c & a  \\
\end{array}} \right) \quad ,
\end{equation}
with $a=2+1/(2m)$, $b=m+2-1/(2m)$, $c=\sqrt{1-1/(4m^2)}$, and $d=m$.
The generality of generating engineered chains for perfect state
transfer is thus clear. On the other hand, it is interesting to note
that in the limit $m\gg1$, one may want to put $a\approx2$,
$b\approx m$ $c\approx1$, and $d=m$. However, since $a \sim O(1)$,
although $m\gg2$, changing $b$ from $m+2$ to $m$ would cause a large
error. This is also confirmed numerically. Therefore, in such a
limit, the requirement of precision is very high. In this sense,
energy spectra that yield more uniform coupling are more desirable
from the engineering point of view.

\subsubsection*{Continuous Systems}

An interesting consequence of Eq. (\ref{eq:spectrum}) can also be
found in infinite dimensional systems. The eigenvalue spectra
allowed by Eq. (\ref{eq:spectrum}) correspond to some canonical
systems such as harmonic well or infinite square well. In those
cases, the necessary criterion for state inversion---namely, the
parity of the eigenstates---is automatically satisfied. For example,
the energy spectrum of an infinite square well is quadratic $E_k
\propto k^2$, for $k=1,2,3,...$ and the eigenfunctions $\psi_k(x)$
have a definite parity $\psi_k(x)=(-1)^{k-1}\psi_k(x)$. One can show
that (also mentioned in \cite{Bransden89}) any single-particle
wave-function $\Psi(x,t)$ at $x$ will be transported (up to a $-$
sign) to $-x$, $\Psi \left( {x,t} \right) = - \Psi \left( { - x,t
{+} \tau } \right)$, for a period of $\tau=2\pi \hbar / E_1$, where
$E_1$ is the ground state energy. This property has been discussed
recently in the literature on fractional wave-function revivals
\cite{Aronstein97}, but its relevance to quantum communication and
its connection to the above general theory linking eigenvalue
spectrum to perfect state transfer has not been appreciated. For
example, one can think of the following strategy for communicating
perfectly through those continuous systems. We can encode the
information of a qudit (not necessarily qubit) to the spin degree of
freedom of a boson or fermion in a state $\Psi(x,t=0)$ which is
initially localized around $x$. At $t=\tau$, the particle will
arrive $\bar x$ and the information can be extracted.

\subsection{Effective Two-Qubit Gates}

\begin{figure}[t]
\centering
\includegraphics[width=8.5cm]{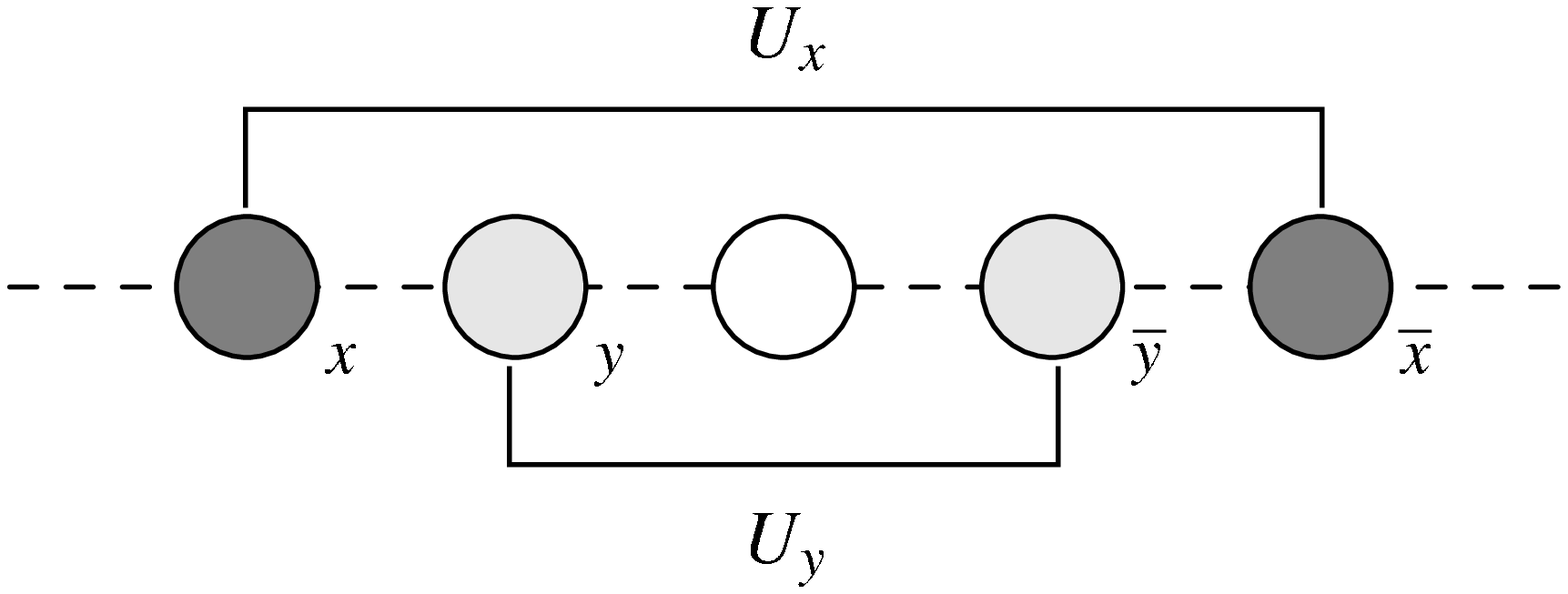}
\caption{In engineered fermionic and bosonic chains, when the
spectrum of the single-particle Hamiltonian in Eq.
(\ref{eq:Tridiagonal}) satisfies the relation (\ref{eq:spectrum}),
effective gates on mirror-conjugate pairs of sites, such as $(x,\bar
x)$ and $(y,\bar y)$, can be constructed by free network
evolution.}\label{fig:gates}
\end{figure}

Two identical fermions (bosons) at sites $x$ and $y$ is described by
the antisymmetrized (symmetrized) product state $\left| \bm {x\,y}
\right\rangle = \left( {1/\sqrt 2 } \right)\left( {\left| \bm x
\right\rangle \! \left| \bm y \right\rangle  \pm \left| \bm y
\right\rangle \! \left| \bm x \right\rangle } \right)$, with $-$
($+$) sign for fermions (bosons). Similarly, the two-particle
eigenstates $\left| \bm {e}_{kl} \right\rangle  = \left( {1/\sqrt 2
} \right)\left( {\left| \bm {e}_k \right\rangle \! \left| \bm {e}_l
\right\rangle \pm \left| \bm {e}_l \right\rangle \! \left| \bm {e}_k
\right\rangle } \right)$ are also antisymmetrized or symmetrized
accordingly. For many fermion excitations, the states are more
convenient to be represented by a Slater determinant. Besides, in
the mapping of spin states to fermionic states, we adopt the
convention \cite{Albanese04,{Clark04}} that the site indices are
arranged in ascending order. Thus, in a spin chain having spin-up
states at $\bm x$ and $\bm y$ but spin-down for all other spins, the
spin state corresponds to the fermionic state $\left| \bm {x\,y}
\right\rangle$ if $x<y$ and $\left| \bm {y\,x} \right\rangle$ if
$y<x$.

By effective gate, we mean the configuration of all the sites after
the network evolution is the same as before, except that the state
of the pair of qubits at $x$ and $\bar x$ is changed according to a
logic gate $U_x$. The simplest way is to choose all other sites to
be empty (or all other spins being the spin-down state for spin
chains). Using similar tricks as before, one can show that
\begin{equation}\label{eq:two excitations}
U(\tau )\left| \bm {x\,y} \right\rangle  = e^{-2i\phi _N } \left|
\bm {\bar x \, \bar{y}} \right\rangle \quad .
\end{equation}
If $y=\bar x$, then there is an extra factor $(-1)$ for fermionic
states but not bosonic states, after exchanging the site indices. We
define a new basis with $\left| {00} \right\rangle \equiv \left| \bm
0 \right\rangle$, $\left| {10} \right\rangle \equiv \left| \bm x
\right\rangle$, $\left| {01} \right\rangle \equiv \left| \bm {\bar
x} \right\rangle$, and $\left| {11} \right\rangle \equiv \left| \bm
{x\,\bar x} \right\rangle$. In this basis, an effective two-qubit
gate $U_{x}$ can be constructed readily for any conjugate pair of
sites $x$ and $\bar x$:
\begin{equation}
U_x  = \left( {\begin{array}{*{20}c}
   1 & 0 & 0 & 0  \\
   0 & 0 & {e^{ - i\phi_N } } & 0  \\
   0 & {e^{ - i\phi_N } } & 0 & 0  \\
   0 & 0 & 0 & { (-1)^\eta e^{-2i\phi_N } }  \\
\end{array}} \right) \quad ,
\end{equation}
where $\eta=1(0)$ for fermions (bosons). The effective gate for a
three-spin chain in \cite{Yung03} is a special case of $U_x$ here.
It is known \cite{Bremner02} that any two-qubit gate that can create
entanglement between two qubits is universal for quantum
computation, when assisted by one-qubit operations. Here we assume
such one-qubit gates are available and we will show that $U_x$ can
create entanglement for fermionic chains (and spin chains) and hence
is universal for quantum computation. For bosonic chains, however,
$U_x$ is not entangling but it allows two-way communication---i.e.,
transfer states from both ends simultaneously.

\subsection{Entanglement Generation and Communication}

The entanglement generation protocols in \cite{Yung03} can now be
generalized. These protocols require minimal spatial and temporal
control of the individual qubits and are also advantageous in that,
after extracting the entangled states at sites $x$ and $\bar x$, the
whole procedures can be repeated by replacing the extracted state
with the corresponding initial states. The configuration of the
intermediate sites or spins will not be changed after each cycle
(except the middle site/spin in protocol 1, which can act as a
trigger of the evolution). Moreover, these protocols can be deployed
for studying the dynamics of entanglement flow \cite{Cubitt04}.

\subsubsection*{Entanglement Generation Protocol 1}

For a linear bosonic or fermionic chain with odd number of sites,
the entanglement generation problem can be mapped to the state
transfer problem. In \cite{Yung03}, only one specific type of
mapping is discussed---namely, the one proportional to $S_x$. Here,
with the enlarged set of choices for the coupling constants, we can
generalize the mapping by including the possibility of nonzero
diagonal coupling terms. For the sake of comparison with protocol~2,
we outline briefly the mapping below.

Suppose the coupling constants still satisfy the symmetry relations
in Eq. (\ref{eq:persymmetric}), we consider a basis consisting of
maximally entangled states $\left| {\tilde j} \right\rangle \equiv
{\textstyle{1 \over {\sqrt 2 }}}\left( {\left| \bm j \right\rangle +
\left| \bm {\bar j} \right\rangle } \right)$ for $j=1,2,...,n-1$,
and a state $\left| {\tilde n} \right\rangle \equiv \left| \bm n
\right\rangle$, where $n = {\textstyle{1\over2}} ({N {+} 1})$ is the
position of the middle site. The Hamiltonian in Eq.
(\ref{eq:Hamiltonian}) acts in this basis as
\begin{equation}
\left( {\begin{array}{*{20}c}
   \lambda_1  & {\omega _{1} } & 0 &  \cdots  & 0  \\
   {\omega _{1} } & \lambda_2 & {\omega _{2} } &  \cdots  & 0  \\
   0 & {\omega _{2} } & \lambda_3 &  \cdots  & 0  \\
    \vdots  &  \vdots  &  \vdots  &  \ddots  & {{\sqrt 2} \omega _{{n{-}1}} }  \\
   0 & 0 & 0 & {{\sqrt 2} \omega _{n{-}1 } } & \lambda_n  \\
\end{array}} \right) \quad ,
\end{equation}
which is also a real and symmetric tridiagonal matrix as $H_N$ (but
the size is about half of it). Suppose the initial state is $\left|
{\tilde n} \right\rangle$---i.e., a single boson or fermion at the
middle site but empty for all other sites. The task of entanglement
generation for the remote pair of sites located at $1$ and $\bar 1$
is the same as to rotate from the unentangled state $\left| {\tilde
n} \right\rangle$ to the entangled state $\left| {\tilde 1}
\right\rangle$. This is equivalent to the state transfer problem we
have discussed and can be solved in exactly the same way.

On the other hand, for linear chains with an even number of sites, a
similar protocol \cite{Yung03} can be used for transferring
entanglement from (now redefined) $\left| {\tilde n} \right\rangle
\equiv {\textstyle{1 \over {\sqrt 2 }}}\left( {\left| \bm n
\right\rangle + \left| \bm {\bar n} \right\rangle } \right)$, where
$n=N/2$, to the remote pair of sites $\left| {\tilde 1}
\right\rangle$. However, this requires the local pair of sites in
the middle to be maximally entangled initially and is therefore an
entanglement transfer protocol.

The two protocols above require the initialization to be made in the
middle of the chains. After the free evolution, the entangled states
are then extracted at the ends of the chains. In situations where we
are allowed to have access only to the pair of sites we want to
entangle, protocol~2, as we shall see next, will be more useful.
However, protocol~1 works for both fermionic and bosonic chains but
protocol~2 is applicable for fermionic chains only.

\subsubsection*{Entanglement Generation Protocol 2}

We now show that any pair of mirror-conjugate sites $x$ and $\bar x$
can be maximally entangled with the application of $U_x$ and the
state initialization at $x$ and $\bar x$ only. For simplicity, all
other sites are set to be empty (or spin-down in applying to spins
chains). First of all, for any normalized pure state of two qubits,
$a\left| {00} \right\rangle + b\left| {01} \right\rangle + c\left|
{01} \right\rangle  + d\left| {11} \right\rangle$, where $\left| a
\right|^2 + \left| b \right|^2 + \left| c \right|^2  + \left| d
\right|^2  = 1$, the concurrence ${\cal C} = 2\left| {ad - bc}
\right|$ is a measure of entanglement \cite{Hill97}. The two sites
are unentangled when ${\cal C}=0$ and maximally entangled when
${\cal C}=1$. Suppose the two sites are initially in a product
state---i.e., $ad=bc$---and all other sites being empty. With the
application of $U_x$, the concurrence becomes $2\left| {a d - \left(
{ - 1} \right)^\eta b c} \right| = 2 \left(1-(-1)^{\eta} \right)
\left| a d \right|$. Consequently, for fermionic chains (with
$\eta{=}1$), the sites $x$ and $\bar x$ can be maximally entangled
from any initial product state with ${ad} = {bc}$ and $\left| {ad}
\right| = {\textstyle{1 \over 4}}$. For example, if the initial
state is $\left| + \right\rangle \left| + \right\rangle$ where
$\left| + \right\rangle  = {\textstyle{1 \over {\sqrt 2 }}} \left(
{\left| 0 \right\rangle + \left| 1 \right\rangle } \right)$, then
the entanglement of the final state can be made explicit by
expressing it in the Schmidt form ${\textstyle{1 \over {\sqrt 2
}}}(| 0 \rangle | {\phi_+} \rangle + e^{ - i\phi _N } | 1 \rangle |
{\phi_-} \rangle)$, where $\left| {\phi _ \pm  } \right\rangle =
{\textstyle{1 \over {\sqrt 2 }}}\left( {\left| 0 \right\rangle \pm
e^{ - i\phi _N } \left| 1 \right\rangle } \right)$ and $\left\langle
{\phi _ + } | {\phi_- } \right\rangle = 0$.

Last, we note that the protocol for generating a class of
multipartite entangled states, called Graph states, suggested in
\cite{Clark04}, can also be extended for the more general
Hamiltonian in (\ref{eq:Hamiltonian}) with various spectrums.

\subsubsection*{Two-way Communication}

For bosonic chains, one can show that the entanglement of any pure
state between sites $x$ and $\bar x$ is invariant after the
application of $U_x$. In fact, the net effect of the free evolution
of the network, with any initial configurations, for a period of
$\tau$ is an inversion of quantum states about the center of the
chain, apart from an extra induced phase $e^{ - i\phi _N }$.
Nonetheless, this implies the possibility of simultaneous transfer
of quantum state from site $x$ to site $\bar x$ and vice versa. Let
us define the protocol more clearly. Suppose Alice and Bob are
sending their states at $x$ and $\bar x$, respectively. We consider
the initial state is in a product state, which can be written in
general as $\left( {a_0 \left| 0 \right\rangle _x + a_1 \left| 1
\right\rangle _x } \right)\left( {b_0 \left| 0 \right\rangle _{\bar
x}  + b_1 \left| 1 \right\rangle _{\bar x} } \right)$, with all
other sites being empty. Applying $U_x$ yields $\left( {b_0 \left| 0
\right\rangle _x + e^{ - i\phi _N } b_1 \left| 1 \right\rangle _x }
\right)\left( {a_0 \left| 0 \right\rangle _{\bar x}  + e^{ - i\phi
_N } a_1 \left| 1 \right\rangle _{\bar x} } \right)$. Therefore,
both states can be sent simultaneously. Interestingly, different
parities can use the same channel, but on different conjugate pair
of sites, at the same time without mutual interference.

\section{Conclusion}
We have demonstrated how to perform quantum state transfer and
construct effective two-qubit gates in engineered networks in which
the coupling constants are determined by the eigenvalue spectrum
satisfying a certain condition. This condition is shown to be both
sufficient and necessary in mirror-symmetrical networks. The
possibility of perfect communication between distant sites of a
single harmonic trap or an infinitely deep square well has been
discussed. The effective gates for fermionic networks, including
spin chains, are entangling and hence can be used for universal
quantum computation. Two entanglement generation schemes are
proposed. The first one works for both fermionic and bosonic chains
but the second one is for the fermionic chains only. Nonetheless,
the bosonic chain allows two-way communication for different pair of
sites simultaneously without mutual interference.



\begin{acknowledgments}
M.H.Y. acknowledge the support of the Croucher Foundation. We thank
S. Benjamin for valuable discussions and D. Burgarth for pointing
out a relevant reference.
\end{acknowledgments}



\end{document}